\documentclass[twocolumn]{aastex63}
\usepackage{times}
\usepackage{graphicx}
\usepackage{tabularx}
\usepackage{amssymb}
\usepackage{amsmath}
\usepackage{pifont}
\usepackage{graphics}
\usepackage{xcolor}
\usepackage{epsfig}
\usepackage{verbatim}
\usepackage{booktabs}
\usepackage{xspace}

\newcommand{\vsinistar}{\ensuremath{v\sin{i_{\star}}}\xspace}
\newcommand{\sinistar}{\ensuremath{\sin{i_{\star}}}\xspace}
\newcommand{\cosistar}{\ensuremath{\cos{i_{\star}}}\xspace}
\newcommand{\msun}{\ensuremath{\,M_\Sun}\xspace}
\newcommand{\rsun}{\ensuremath{\,R_\Sun}\xspace}

\newcommand{\mj}{\ensuremath{\,M_{\rm J}}\xspace}

\newcommand{\kms}{\ensuremath{\rm km\ s^{-1}}\xspace}
\newcommand{\ms}{\ensuremath{\rm m\ s^{-1}}\xspace}
\newcommand{\mstar}{\ensuremath{M_{\star}}\xspace}
\newcommand{\rstar}{\ensuremath{R_{\star}}\xspace}

\newcommand{\ktwo}{\emph{K2}\xspace}
\newcommand{\spitzer}{\emph{Spitzer}\xspace}
\newcommand{\planet}{V1298~Tau~b\xspace}
\newcommand{\starname}{V1298~Tau\xspace}
\newcommand{\prot}{$P_\mathrm{rot}$\xspace}
\newcommand{\gaia}{\textit{Gaia}}

\newcommand{\lambadfinal}{$\lambda = 4_{-10}^{+7}$~deg\xspace}
\newcommand{\psifinal}{$\psi = 8_{-7}^{+4}$~deg\xspace}
\newcommand{\imutfinal}{$i_{\mathrm{mut}}=0^{\circ} \pm 19^{\circ}$\xspace}
\newcommand{\vsinifinal}{\vsinistar$ = 24.87^{+0.19}_{-0.21}$ \kms\xspace}
\newcommand{\agefinal}{$28\pm4$\,Myr\xspace}

\begin{document}
\title{An Aligned Orbit for the Young Planet V1298 Tau b}

\author[0000-0002-5099-8185]{Marshall C. Johnson} 
\affiliation{Las Cumbres Observatory, 6740 Cortona Drive, Suite 102, Goleta, CA 93117, USA}
\affiliation{Department of Astronomy, The Ohio State University, 4055 McPherson Laboratory, 140 West 18$^{\mathrm{th}}$ Ave., Columbus, OH 43210 USA}
\email{johnson.7240@osu.edu}

\author[0000-0001-6534-6246]{Trevor J.\ David} %HIRES observations, RVRM modeling
\affil{Center for Computational Astrophysics, Flatiron Institute, 162 5$^{\mathrm{th}}$ Ave. New York, NY 10010, USA}
\affil{Department of Astrophysics, American Museum of Natural History, 200 Central Park West, New York, NY 10024, USA}

\author[0000-0003-0967-2893]{Erik A.\ Petigura} %HIRES observations
\affil{Department of Physics and Astronomy, University of California, Los Angeles, 430 Portola Plaza, Box 951547, Los Angeles, CA 90095, USA}

\author[0000-0002-0531-1073]{Howard T.\ Isaacson} %confirmed %HIRES reduction, a lot of work looking into the 15 km/s offset between the blue and red chips
\affil{Department of Astronomy, University of California, Berkeley, 501 Campbell Hall \#3411, Berkeley, CA 94720, USA}

\author{Judah Van Zandt} %HIRES transit observations; confirmed
\affil{Department of Physics and Astronomy, University of California, Los Angeles, 430 Portola Plaza, Box 951547, Los Angeles, CA 90095, USA}

\author{Ilya Ilyin} %PEPSI observations and reduction
\affiliation{Leibniz-Institut f\"ur Astrophysik Potsdam (AIP), An der Sternwarte 16, D-14482 Potsdam, Germany}

\author{Klaus Strassmeier} %PEPSI team; confirmed
\affiliation{Leibniz-Institut f\"ur Astrophysik Potsdam (AIP), An der Sternwarte 16, D-14482 Potsdam, Germany}

\author{Matthias Mallonn} %PEPSI team; confirmed
\affiliation{Leibniz-Institut f\"ur Astrophysik Potsdam (AIP), An der Sternwarte 16, D-14482 Potsdam, Germany}

\author{George Zhou} %TRES observations, confirmed
\affiliation{Centre for Astrophysics, University of Southern Queensland, USQ Toowoomba, West Street, QLD 4350, Australia}

\author[0000-0003-3654-1602]{Andrew W. Mann} %proposal team, revised age calculation, has a very cute cat. 
\affiliation{Department of Physics and Astronomy, University of North Carolina at Chapel Hill, 120 E. Cameron Ave., Phillips Hall CB3255, Chapel Hill, NC 27599, USA}

\author[0000-0002-4881-3620]{John~H.~Livingston} %Spitzer data, confirmed
\affiliation{Department of Astronomy, University of Tokyo, 7-3-1 Hongo, Bunkyo-ku, Tokyo 113-0033, Japan}

\author[0000-0002-0296-3826]{Rodrigo Luger} %Derivation of mutual inclination formula. Some guidance on RVRM modeling; confirmed
\affil{Center for Computational Astrophysics, Flatiron Institute, 162 5$^{\mathrm{th}}$ Ave. New York, NY 10010, USA}

%Other HIRES observers
\author[0000-0002-8958-0683]{Fei Dai} %confirmed; also suggested spot analysis
\affiliation{Division of Geological and Planetary Sciences, 1200 E California Blvd, Pasadena, CA, 91125, USA}

\author{Lauren M. Weiss} %confirmed
\affiliation{Department of Physics, University of Notre Dame, 225 Nieuwland Science Hall, Notre Dame, IN 46556, USA}

\author[0000-0003-4603-556X]{Teo Mo\v{c}nik} %confirmed
\affiliation{Gemini Observatory/NSF's NOIRLab, 670 N. A'ohoku Place, Hilo, HI 96720, USA}

\author{Steven Giacalone} %confirmed
\affil{Department of Astronomy, University of California, Berkeley, 501 Campbell Hall \#3411, Berkeley, CA 94720, USA}

\author[0000-0002-0139-4756]{Michelle L. Hill} %confirmed
\affiliation{Department of Earth and Planetary Sciences, University of California Riverside, 900 University Ave, Riverside, CA 92521, USA}

\author[0000-0002-7670-670X]{Malena Rice} %confirmed
\affil{Department of Astronomy, Yale University, New Haven, CT 06511, USA}

\author{Sarah Blunt} %confirmed
\affil{Astronomy Department, California Institute of Technology, 1200 East California Blvd, Pasadena CA 91125, USA}

\author{Ryan Rubenzahl} %confirmed
\affil{Astronomy Department, California Institute of Technology, 1200 East California Blvd, Pasadena CA 91125, USA}

\author[0000-0002-4297-5506]{Paul A. Dalba} %confirmed
\altaffiliation{NSF Astronomy and Astrophysics Postdoctoral Fellow}
\affiliation{Department of Astronomy and Astrophysics, University of California, Santa Cruz, CA 95064, USA}
\affiliation{Department of Earth and Planetary Sciences, University of California Riverside, 900 University Ave, Riverside, CA 92521, USA}

%%% Other TRES Authors

\author{Gilbert A. Esquerdo} %esquerdo@psi.edu %confirmed
\affiliation{Center for Astrophysics \textbar{} Harvard \& Smithsonian, 60 Garden St., Cambridge, MA 02138, USA.}

\author{Perry Berlind}  %confirmed %pberlind@cfa.harvard.edu
\affiliation{Center for Astrophysics \textbar{} Harvard \& Smithsonian, 60 Garden St., Cambridge, MA 02138, USA.}

\author{Michael L. Calkins} %confirmed %mcalkins@cfa.harvard.edu
\affiliation{Center for Astrophysics \textbar{} Harvard \& Smithsonian, 60 Garden St., Cambridge, MA 02138, USA.}

\author[0000-0002-9328-5652]{Daniel Foreman-Mackey} %Some guidance on RVRM modeling
\affil{Center for Computational Astrophysics, Flatiron Institute, 162 5$^{\mathrm{th}}$ Ave. New York, NY 10010, USA}

\shorttitle{V1298 Tau \MakeLowercase{b} Is Aligned}

\begin{abstract}
The alignment of planetary orbits with respect to the stellar rotation preserves information on their dynamical histories. Measuring this angle for young planets help illuminate the mechanisms that create misaligned orbits for older planets, as different processes could operate over timescales ranging from a few Myr to a Gyr. We present spectroscopic transit observations of the young exoplanet V1298 Tau b; we update the age of V1298 Tau to be \agefinal based on Gaia EDR3 measurements. We observed a partial transit with Keck/HIRES and LBT/PEPSI, and detected the radial velocity anomaly due to the Rossiter-McLaughlin effect. V1298 Tau~b has a prograde, well-aligned orbit, with \lambadfinal. By combining the spectroscopically-measured \vsinistar\ and the photometrically-measured rotation period of the host star we also find that the orbit is aligned in 3D, \psifinal. 
 Finally, we combine our obliquity constraints with a previous measurement for the interior planet V1298 Tau c to constrain the mutual inclination between the two planets to be \imutfinal. 
 This measurements adds to the growing number of well-aligned planets at young ages, hinting that misalignments may be generated over timescales of longer than tens of Myr. The number of measurements, however, is still small, and this population may not be representative of the older planets that have been observed to date.
 We also present the derivation of the relationship between $i_{\mathrm{mut}}$, $\lambda$, and $i$ for two planets.
\end{abstract}


\section{Introduction}
\label{sec:Intro}
Planets are formed in circumstellar disks, which themselves form as a result of angular momentum conservation during the collapse of protostellar cores. The angular momentum vectors of stars and protoplanetary disks should therefore be aligned at the start of the planet formation process. Even the modest tilt of six degrees between the Sun's spin axis and the mean angular momentum plane of the solar system is seen as a curiosity demanding a physical explanation \citep{Kuiper51}. Proposed mechanisms capable of producing this tilt include asymmetric infall or torques from mass concentrations in the collapsing core \citep{Tremaine91}, encounters with other stars in the birth cluster \citep{Heller93}, the presence of an undiscovered planet in the Solar System \citep{Bailey16, Gomes17}, and a misalignment between the solar spin and mean solar wind axes \citep{Spalding19}. 

In contrast with the mildly misaligned solar system, exoplanets are routinely found on well-aligned, grossly misaligned, polar, and even retrograde orbits \citep{DawsonJohnson18}.  
How and when stellar spin-orbit misalignments arise are open questions. Some spin-orbit misalignments may be primordial, the result of, for example, torquing of the protoplanetary disk by a distant companion, nearby star, or other gas aggregation in the birth cluster \citep{Heller93, Thies11, Batygin11, Batygin20, Batygin12, SpaldingBatygin14}. Early stage misalignments may also arise from asymmetric, variable, and turbulent accretion \citep{Tremaine91, Bate10, Fielding15}, magnetic star-disk interactions \citep{Lai11, FoucartLai11, BatyginAdams13, Lai14, SpaldingBatygin14, SpaldingBatygin15}, or planet-disk interactions \citep{Millholland19, SuLai20}. Still other misalignments may emerge long after the planet has formed, as a result of dynamical interactions with other planets or a companion star \citep{FabryckyTremaine07, Wu07, Naoz11, Storch14}. 

By measuring exoplanet obliquities for systems of varying ages it may be possible to determine when spin-orbit misalignments typically arise, which in turn can help to determine the dominant mechanism(s) producing misalignments. If misalignments are predominantly generated by disk tilting, then misaligned orbits should be seen for planets of all ages. If, on the other hand, misalignments are typically generated by slower mechanisms like secular chaos or the Kozai-Lidov mechanism, then misaligned orbits should begin to appear at ages of order hundreds of Myr. Planet-planet scattering should operate on intermediate timescales of tens of Myr. 
Obliquity measurements for young exoplanets are particularly intriguing as those systems should more closely reflect the initial conditions of a given planetary system. In particular, some planets which start off misaligned may become realigned over time due to tidal interactions with the host star \citep{Albrecht12}. 

Here we report on the obliquity of V1298 Tau~b (aka EPIC 210818897~b or K2-309~b), a moderately-irradiated (24 day period), Jupiter-sized exoplanet transiting a young solar analog with an age of 20--30 Myr \citep{David19}.  The planet is one of at least four transiting planets in the system \citep{David19b}, and \cite{SuarezMascareno21} recently measured a mass of 0.64 \mj\ for this planet. Given the presence of multiple transiting planets, a low mutual inclination between the planets is implied, but there is no requirement that the stellar spin axis is aligned with the orbital angular momentum vectors of the planets.  

Our analysis is based on contemporaneous high-resolution time-series spectroscopy acquired with the Keck/HIRES, LBT/PEPSI, and FLWO/TRES spectrographs, covering a single transit event. We also include transit photometry from the \textit{K2} mission and \textit{Spitzer} Space telescope %, and LCO 
to accurately constrain the planet's ephemeris and other transit parameters. We use both the Rossiter-McLaughlin (RM) and Doppler Tomography (DT) techniques to attempt to infer the obliquity of V1298 Tau b. 

The remainder of the paper is organized as follows: In \S~\ref{sec:Obs} we describe our spectroscopic observations and auxiliary time series photometry. We describe our analysis methods in \S~\ref{sec:Analysis} and discuss our findings in the context of other exoplanet studies in \S~\ref{sec:Discussion}. Finally, we present our conclusions in \S~\ref{sec:Conclusions}.

\section{Observations}
\label{sec:Obs}

\subsection{Spectroscopic Observations}
Below we describe the spectroscopic observations which form the basis of our analysis. The radial velocities derived from these observations are listed in Table~\ref{tab:rvs}. 

\subsubsection{Keck/HIRES}
We observed a partial transit of V1298 Tau b on 2019 Oct. 24 UT with the 10m Keck I telescope on Maunakea, Hawaii, USA, and the HIRES spectrograph \citep{Vogt94}. HIRES is a cross-dispersed \'echelle spectrograph. We used the C2 decker, with a slit of 0.861'' width and 14.0'' height, yielding a resolving power of $R\sim60,000$ and the ability to subtract background light including scattered moonlight and night sky emission lines. The spectra span from 3646 to 7984~\AA, although the velocities are derived from the 5000--6000~\AA\ region . 

Our observations began 5.7 hours before transit and continued through the first 75\% of the transit, terminating at twilight. These were interrupted only by obtaining a template spectrum of V1298 Tau during the pre-transit phase (which is included in our Doppler tomographic time series but not our RV time series), and a brief interruption during transit while the target passed through the zenith and the telescope tracking could not keep up. During this time we obtained 54 spectra, two of which were template spectra without the I$_2$ cell in the light path, and the remaining 52 of which used the I$_2$ cell. Radial velocities (RVs) were derived from the HIRES spectra using the California Planet Survey (CPS) pipeline described in \cite{Howard10CPS}. We included an additional 8 HIRES RVs acquired within 20 days of the transit night in order to train a Gaussian process, as described in \S\ref{subsec:RVRM}. These RVs were acquired as part of an ongoing program and derived in an identical fashion to those acquired during the transit night.

\subsubsection{LBT/PEPSI}

We observed the same transit with the Potsdam Echelle Polarimetric and Spectroscopic Instrument \citep[PEPSI,][]{Strassmeier15} at the 2 $\times$ 8.4 m Large Binocular Telescope (LBT) on Mount Graham, Arizona, USA. We used  PEPSI's $R=120,000$ mode and cross dispersers III and V (covering 4800-5441 \AA\ and 6278-7419 \AA). We acquired 24 spectra (with exposures of 900 seconds). %Owing to the more easterly location we were able to gather
We began observing 4.4 hours before ingress, but were only able to observe the first 25\% of the transit before morning twilight due to LBT's more easterly location with respect to Keck.

The data reduction was performed with the Spectroscopic Data Systems (SDS) pipeline, adapted to the PEPSI data calibration flow and image specific content. It was based upon the pipeline described in \cite{Ilyin2000} and a recent description is given in \cite{Strassmeier18}.

The specific steps of image processing included bias subtraction and variance estimation of the source images, super-master flat field correction for the CCD spatial noise, \'echelle orders definition from the tracing flats, scattered light subtraction, and a wavelength solution for the ThAr images. The pipeline then performed optimal extraction of image slicers and cosmic ray spikes elimination of the target image, wavelength calibration and merging slices in each order, and normalization to the master flat field spectrum to remove CCD fringes and blaze function. Finally, it performed a global 2D fit to the continuum of the normalized image, and rectification of all spectral orders in the image to a 1D spectrum for a given cross-disperser.

The spectra from two sides of the telescope were averaged with weights into one spectrum and corrected for the barycentric velocity of the Solar system. The wavelength scale was preserved for each pixel as given by the wavelength solution without rebinning.  The wavelength solution used about 3000 ThAr lines and had the error of the fit at the image center of 4~\ms.

The time series spectra were cross-correlated in the blue and red arms separately with respect to the first three averaged spectra which was used as the reference spectrum. A pre-selected spectral region in the red with low telluric contamination was used for cross-correlation. It used a weighted linear regression fit of the reference spectrum to the observed one for every radial velocity offset of the re-sampled reference spectrum. The accuracy of the offset was defined by the curvature of the cross-correlation peak and $\chi^2_\nu$ of the regression fit.

\subsubsection{FLWO 1.5m/TRES}

\begin{figure}
    \centering
    \includegraphics[width=\linewidth]{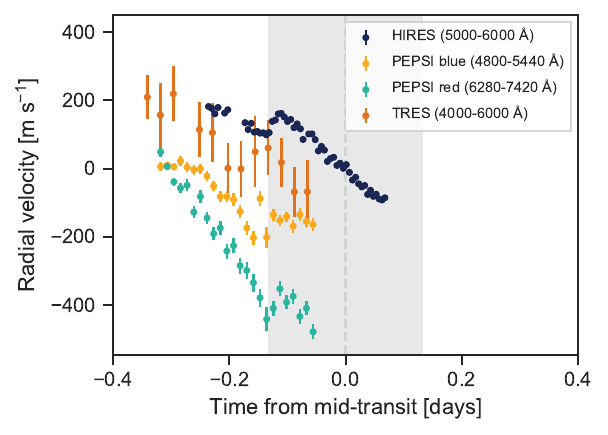}
    \caption{Comparison of time series RVs acquired by different spectrographs on the night of the transit of V1298 Tau b. Offsets and trends of varying slopes between instruments are apparent. The gray shaded band indicates the predicted transit window. The chromaticity of the RV trend is apparent from the PEPSI data. We note that wavelength ranges indicated are the approximate regions from which radial velocity information is extracted, and not necessarily the complete wavelength ranges of the spectra.}
    \label{fig:instruments}
\end{figure}

We obtained 21 epochs of observations of V1298 Tau using the Tillinghast Reflector Echelle Spectrograph \citep[TRES,][]{Furesz:2008} on the 1.5\,m telescope at the Fred Lawrence Whipple Observatory, Arizona, USA. Nine epochs were obtained over the two weeks prior to the transit event on 24 October 2019 UT, and 12 were obtained on the night of the transit. TRES is a fiber fed echelle with a spectral resolving power of $R\approx44,000$ over the wavelength range of $3850-9100$\,\AA. Each epoch consists of three consecutive exposures that are median combined to reduce the impact of cosmic rays, and are bracketed by a set of Thorium-Argon hollow cathode lamp exposures that provide the wavelength solution to the spectrum. The spectra are calibrated and extracted as per \citet{2012Natur.486..375B}. To measure the radial velocities, we derived line profiles from each spectrum via an least squares deconvolution analysis \citep{Donati97}, and fit the broadening profiles with a joint kernel describing the effects of the Doppler shift, rotational, macroturbulent, and instrumental broadening. The derived velocities are presented in Table~\ref{tab:rvs} and Figure~\ref{fig:instruments}. 

\begin{deluxetable}{lccc}
\tabletypesize{\footnotesize}
\tablecolumns{4}
\tablewidth{0pt}
\tablecaption{Radial velocities of V1298 Tau. \label{tab:rvs}}
\tablehead{
\colhead{Date} &
\colhead{RV} &
\colhead{Error} &
\colhead{Instrument} \\
\colhead{(BJD)} & 
\colhead{(\ms)} & 
\colhead{(\ms)} & 
\colhead{}
}
\startdata
2458780.853237 & 182.2 & 6.2 & HIRES \\
2458780.858724 & 176.9 & 6.1 & HIRES \\
2458780.864210 & 160.9 & 6.2 & HIRES \\
2458780.869662 & 179.4 & 6.0 & HIRES \\
2458780.881040 & 163.0 & 5.9 & HIRES \\
\enddata
\tablecomments{Table 1 is published in its entirety in the machine-readable format. A portion is shown here for guidance regarding its form and content.}
\end{deluxetable}

\subsection{Photometric Observations}

In addition to our spectroscopic observations, we used two photometric datasets in our fits in order to accurately derive the planetary ephemeris. We describe these data below.

\subsubsection{K2}
The \textit{Kepler} space telescope observed V1298 Tau between 2015 Feb 7 and 2015 Apr 23	UT during Campaign 4 of its extended \textit{K2} mission \citep{Howell14}. We used \textit{K2} photometry in conjunction with our RVs to determine accurate and robust uncertainties on the sky-projected obliquity and related transit parameters. We used the flattened light curve from \citet{David19b}, which in turn used the \texttt{EVEREST 2.0} pipeline \citep{Luger16, Luger18} to optimize the photometric aperture and correct the light curve for instrumental systematics using pixel-level decorrelation.

\subsubsection{Spitzer}
A partial transit of V1298 Tau b was observed in the IRAC2 (4.5~\micron) channel on the \textit{Spitzer} space telescope on 2019 June  1 UT (Livingston et al., in prep.). We direct the reader to that paper for details on the {\it Spitzer} data processing procedures. In this work we are only concerned with accurately measuring the planet's ephemeris and other transit parameters, and we use the corrected light curve presented by those authors. 

From the \citet{David19b} ephemeris, based only on \textit{K2} data from 2015, the predicted times (in BJD) of ingress, mid-transit, and egress on the night of our observations were $\approx$ 2458780.82665, 2458780.96040, and 2458781.09415, respectively. Including the \textit{Spitzer} data instead constrained these times to be 2458780.95530, 2458781.08905, and 2458781.22280 (again in BJD). Therefore, including the \textit{Spitzer} constraints shifts the transit 3 hours later on our night of observations.

\section{Analysis}
\label{sec:Analysis}

\subsection{Reassessment of V1298 Tau's Age}
\label{sec:age}

With the recent arrival of \gaia\ Data Early Data Release 3 \citep[EDR3;][]{Gaia16,GaiaEDR3}, it should be possible to derive a more precise age measurement than was possible in \cite{David19}. To this end, we used the Group 29 membership list from \citet{Luhman18} with updated photometry and parallaxes from \gaia\ EDR3. We removed stars with RUWE$>1.2$, as these are more likely to be binaries \citep{Ziegler2020}. We compared the resulting color-magnitude diagram to predictions from the PARSECv1.2S models \citep{PARSEC} assuming Solar abundance. To account for both binaries and non-member interlopers, we perform the comparison using a mixture model following \citet{HoggRecipes} and described in more detail in Mann et al (in prep). To briefly summarize, we compared photometry for candidate members to a two-component model; the first component is single-star single-aged population drawn from the isochrone (for a given $E(B-V)$ and age), and the second component an outlier population described by an offset from the first component ($Y_B$) and a variance around that offset ($V_B$). There were two additional free parameters, one to capture underestimated uncertainties, differential reddening, and small age spreads ($f$), and one for the amplitude of the second population ($P_B$) such that a pure single-star single-age population should have $P_B=0$. All parameters evolved under uniform priors. We performed the comparison in an MCMC framework using \texttt{emcee} with 50 walkers and 100,000 steps following a burn-in of 10,000 steps. 

\begin{figure}
\centering
\includegraphics[width=\columnwidth]{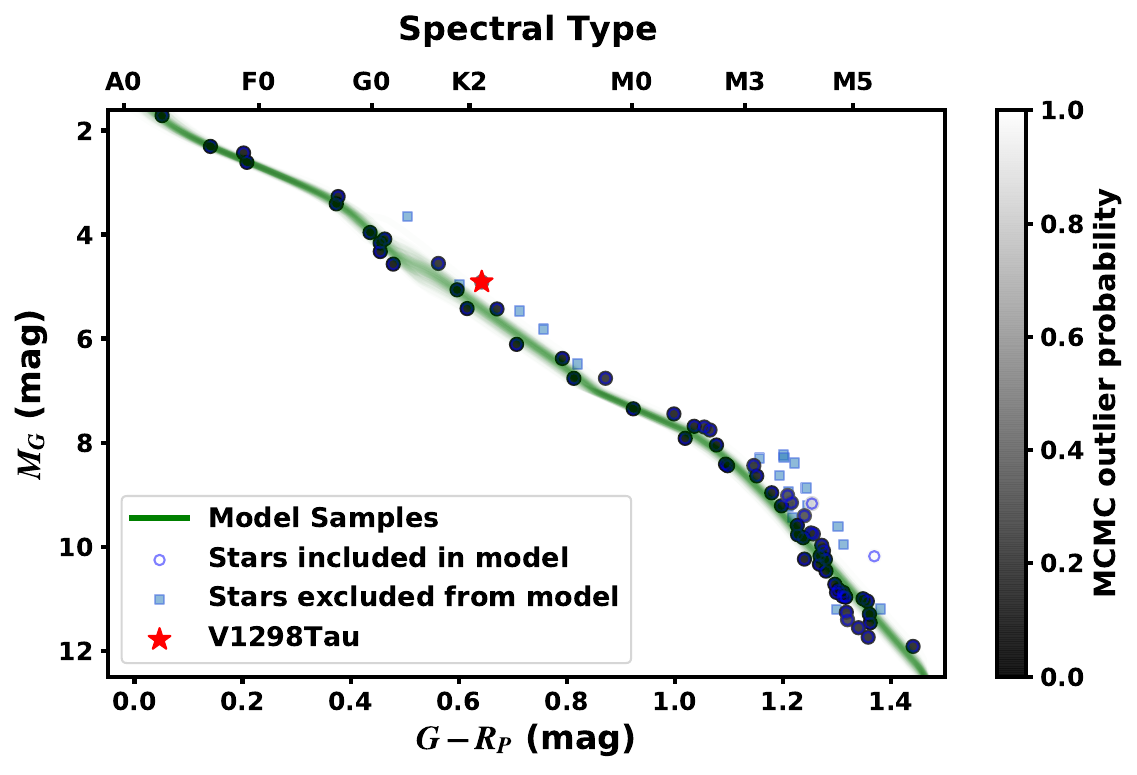}
\caption{Example comparison of likely members of Group 29 to a model isochrone from PARSEC. Each point is a star from \cite{Luhman18}, shaded by their probability of being part of the outlier distribution (the second model in the mixture). Squares are targets that were excluded from the fit due to a high RUWE, although including them made no difference (they were captured as outliers in then mixture). The green lines are 100 random model samples from the MCMC. V\,1298\,Tau is denoted as a red star. Approximate spectral types are listed on the top axis.
}
\label{fig:cmd}
\end{figure}

We show the resulting fit in Figure~\ref{fig:cmd} (see also Figure~\ref{fig:corner-cmd} for a corner plot), which yielded an age of $30\pm3$\,Myr. This uncertainty does not fully account for systematic errors in the models or uncertainties arising from the sample selection. The latter may be especially important, as V\,1298\,Tau is not on the membership list from \citet{Luhman18} and the greater Taurus-Aurgia association is likely a mix of many populations \citep{2021arXiv210513370K}. To test for model errors, we repeated the process using models from the Dartmouth Stellar Evolution Program \citep[DSEP;][]{Dotter2008} with magnetic enhancement \citep{Feiden2012b}. To test for problems with the sample selection, we repeated the comparison with each model grid using a sample of stars with tangential velocities within 2\kms\ and locations within 25\,pc of V\,1298\,Tau selected using the \texttt{FriendFinder}\footnote{\url{https://github.com/adamkraus/Comove}} algorithm \citep{Tofflemire2020}. \texttt{FriendFinder} stars were closer (kinematically and spatially) to V 1298 Tau, and hence more likely to be from the same initial cloud, but this selection also contained some non-member interlopers (evident from a low color-magnitude diagram [CMD] position). Fortunately, our mixture model simply models these stars as outliers and they have a small or negligible impact on the final age.

For all four fits, the estimated age was between 25 and 32\,Myr. Based on this, we adopt a conservative age of \agefinal, consistent with earlier 20--30\,Myr assessments \citep{Luhman18,David19b}. Notably, V\,1298\,Tau sits above the CMD in all fits, although it is consistent with the scatter seen around the model sequence for other members. 

\subsection{Stellar Rotational Velocity}
\label{sec:vsini}

Obtaining an accurate measurement of the spin-orbit misalignment from the Rossiter-McLaughlin effect depends critically upon having an accurate measurement of the projected stellar rotational velocity \vsinistar. V1298 Tau is sufficiently rapidly rotating that we can spectroscopically resolve the rotationally broadened line profile and directly measure \vsinistar. 

We use the higher-resolution PEPSI data for this purpose. We extracted the average line profile from the PEPSI spectra as described in \S\ref{sec:DT}. We fit these data using the \texttt{misttborn} package\footnote{\url{https://github.com/captain-exoplanet/misttborn}} \citep[e.g.,][]{DT3,Dholakia19}. We produced a model rotationally broadened line profile as described in \cite{Kepler13Ab}; briefly, we assume that the line profile from each stellar surface element is a Gaussian appropriately Doppler shifted according to the RV of the surface element assuming solid body rotation. We assumed a quadratic limb-darkening law, and used the triangular sampling method of \cite{Kipping13}. We then numerically integrate over a Cartesian grid on the surface of the star. \texttt{misttborn} uses the affine-invariant Markov chain Monte Carlo package \texttt{emcee} \citep{ForemanMackey13} to produce posterior distributions for the parameters. 

We present the results from this fit in Table~\ref{table:lineprofile}. Our value of \vsinifinal is consistent with but more precise than previous measurements \citep{David19}. We adopt this value for the remainder of the analyses in this paper. 

\begin{table} 
\label{table:lineprofile}
\caption{Results of Line Profile Fit} 
\begin{tabular}{lc} 
\hline 
\hline 
Parameter & Value \\ 
\hline 
Measured Parameters \\ 
$v\sin i_{\star}$ (km s$^{-1}$) & $24.87^{+0.19}_{-0.21}$ \\ 
$v_{\mathrm{int}}$ (km s$^{-1}$) & $6.20^{+0.16}_{-0.17}$ \\ 
$q_{1}$ & $0.351^{+0.104}_{-0.099}$ \\ 
$q_{2}$ & $0.166^{+0.079}_{-0.067}$ \\ 
%linecenter1 & $-0.184^{+0.015}_{-0.2}$ \\ 
\hline 
Derived Parameters \\ 
$g_{1}$ & $0.193^{+0.071}_{-0.074}$ \\ 
$g_{2}$ & $0.4^{+0.12}_{-0.13}$ \\ 
\hline 
\end{tabular} 
\tablecomments{Parameters derived from the fit to the line profile of V1298 Tau. $v_{\mathrm{int}}$ is the intrinsic width of a Gaussian line of each stellar surface element. $q$ are the triangularly-sampled limb darkening parameters per \cite{Kipping13}, while $g$ are the corresponding quadratic limb-darkening parameters.}
\end{table}

\subsection{Radial Velocity Rossiter-McLaughlin Analysis}
\label{subsec:RVRM}

We show our radial velocities on the night of the transit in Fig.~\ref{fig:instruments}. It is apparent from this figure that we detect the transit with both HIRES and PEPSI, and that the orbit is prograde; there is a sharp positive deviation in the RVs from both instruments during ingress. However, there is also a large out-of-transit slope in the RVs, which varies from instrument to instrument, and, indeed, even between the RVs derived from the red and blue arms of PEPSI. We ascribe this RV trend to starspots rotating across the face of V1298 Tau. Starspots are chromatic, and will therefore have different influences on RVs measured at different wavelengths.

We jointly modeled the RVs and time series photometry using the \texttt{starry 1.1.2} \citep{exoplanet:luger19} and \texttt{exoplanet 0.5.1} \citep{exoplanet:exoplanet, exoplanet:joss} packages. Stellar surface brightness and velocity field variations are modeled as an expansion of spherical harmonics in \texttt{starry}. These variations, and their derivatives, are computed analytically so \texttt{starry} models can be easily incorporated in gradient-based optimization and inference procedures such as Hamiltonian Monte Carlo (HMC), No U-Turns Sampling (NUTS), and variational inference. Detailed descriptions of the application of \texttt{starry} to Rossiter-McLaughlin measurements are provided in \cite{Bedell19} and \cite{Montet20}. 

The inferred sky-projected obliquity is sensitive to the underlying RV trend. This is particularly true for the data set in consideration as the RVs only cover a partial transit (complete transits are rarely observable from a single site due to the long orbital period and transit duration). In an idealized, perfectly quiet star the RV trend is due to the reflex motion of the star due to all orbiting companions. In the present case, \starname has prominent starspots which induce apparent RV variations that are more than an order of magnitude larger ($\sim$200~\ms\ in semi-amplitude) than the expected Doppler semi-amplitudes of any of the known planets in the system. Consequently, in the modeling that follows we neglect reflex motions due to the planets. 

In order to investigate the sensitivity of the inferred obliquity on the assumed RV trend we explored a number of different approaches to modeling the underlying RV variations: (1) a linear trend, (2) a quadratic trend, and (3) a quasi-periodic Gaussian process (GP). For the Gaussian process kernel we selected the quasi-periodic Rotation Term kernel implemented in \texttt{exoplanet}, a combination of two stochastically driven simple harmonic oscillators which has been shown to successfully reproduce stellar variability \citep{Angus2018}. Further details about this kernel and its associated hyperparameters are presented in \citet{David19b}. We note that in the GP modeling of the RVs, we chose not to use the PEPSI red data as it showed a much steeper out-of-transit slope than the PEPSI blue, HIRES, and TRES data and thus was not well-described by our model which adopts a single GP amplitude across all instruments.

For the linear and quadratic trend models we jointly fit only the Keck/HIRES RVs (as these are the most precise and present the clearest detection of the RM anomaly) along with the flattened \textit{K2} and \textit{Spitzer} light curves. For the GP, we included RVs acquired within 20 days of the RM observations in order to train the GP. We included constant offsets and jitter parameters for each individual instrument in the RV data set as well as for the \textit{K2} and \textit{Spitzer} time series. Since starspot crossings affect the transit depths of the K2 light curve, we allowed the jitter parameter for that data set to scale linearly with the transit depth in order to account for the excess noise in transit. In all models we assumed a linear ephemeris (i.e. we did not include transit timing variations) across the RV, \spitzer, and \ktwo data sets; we were able to obtain a good fit to all five individual transit events using this assumption. We also assumed a single $R_P/R_*$ value for all data sets. \ktwo and the RV spectrographs use broadband optical passbands, and the \textit{Spitzer} transit depth has been found to be consistent with the \textit{K2} transit depth at the 1$\sigma$ level (Livingston et al., in prep.). We also neglect eccentricity in the orbit of planet b as transit modeling and orbit-crossing constraints suggest it is small, $<$0.3 \citep{David19b}.

The rotation period of V1298 Tau was precisely measured from the \ktwo light curve via Gaussian process regression \citep[$P_\mathrm{rot} = 2.87 \pm 0.02$~d,][]{David19b}. We incorporate this information into our model by placing a Gaussian prior on the inferred equatorial rotation period, where that quantity is derived as $P_\mathrm{rot,eq} = 2 \pi R_\star / v_\mathrm{eq}$, and $R_\star$ is a directly sampled parameter. The equatorial velocity, $v_\mathrm{eq}$ is derived from \vsinistar (a free parameter) and \sinistar (derived from \cosistar, which is sampled from a uniform distribution between 0 and 1) using $v_\mathrm{eq}=v\sin{i_\star}/\sin{i_\star}$. 

We note that a limitation of this model is that it does not explicitly account for surface differential rotation, as we do not know what latitude(s) gave rise to the K2 brightness modulations. Differential rotation may be as high as $\sim$0.4 rad d$^\mathrm{-1}$ for young stars \citep[e.g.][]{Marsden06, Marsden11b}. For reference, differential rotation of 0.2 rad d$^\mathrm{-1}$ in \starname would result in a pole-to-equator difference of 0.2--0.3~d in the rotation period, an order of magnitude larger than the uncertainty reported from the Gaussian process regression. To account for the uncertainty introduced by differential rotation, for each model we tested two different \prot priors in the HMC sampling: Gaussian priors centered on the photometrically determined rotation period with widths of 0.2~d and 0.02~d (the concentrated prior). 

Notably, the constraints on the rotation period, \vsinistar, and \rstar allow the stellar inclination to be inferred which,  combined with the planet's orbital inclination and sky-projected obliquity, allows for derivation of the spin-orbit angle \citep[see Equation 9 of][]{FabryckyWinn09}.

To infer robust uncertainties on the parameters in our models we used the NUTS sampler as implemented in \texttt{PyMC3 3.11.4} \citep{exoplanet:pymc3}. After finding the maximum a posteriori (MAP) model we initiated the sampler using 5000 tuning steps, a target acceptance fraction of 95\%, and 2 chains each with 2500 draws from the posterior. Convergence was assessed using the Gelman-Rubin statistic \citep{GelmanRubin1992}. 

\begin{figure*}
    \centering
    \includegraphics[width=0.33\linewidth]{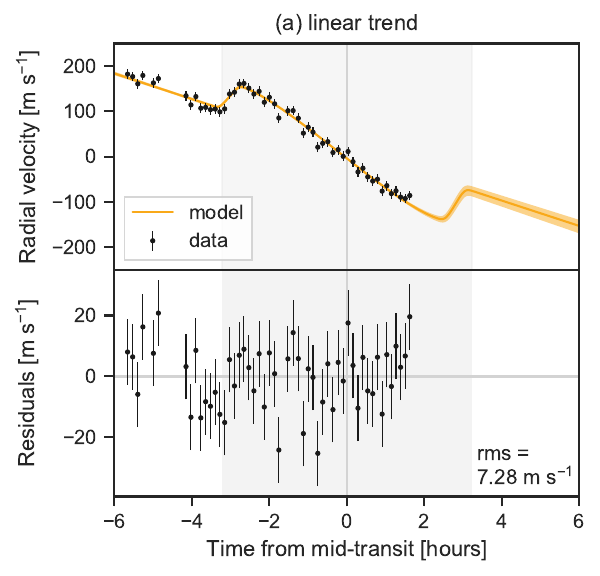}
    \includegraphics[width=0.33\linewidth]{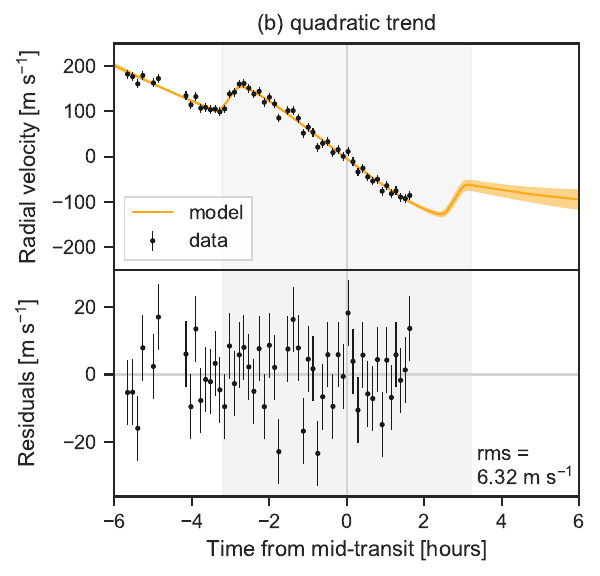}
    \includegraphics[width=0.33\linewidth]{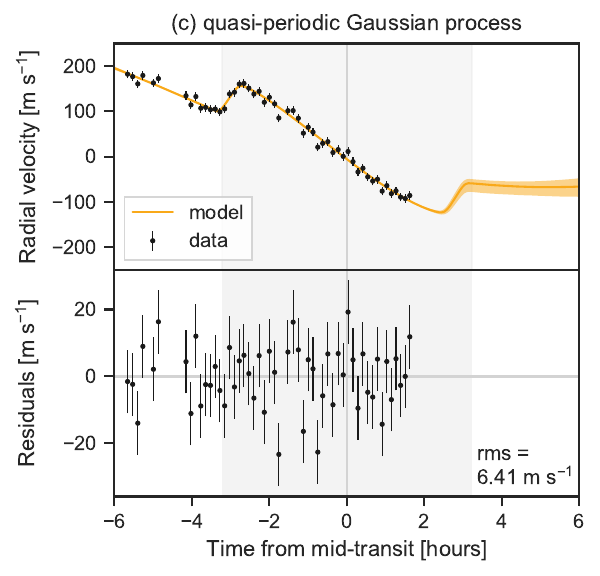} \\
    \includegraphics[width=0.66\linewidth]{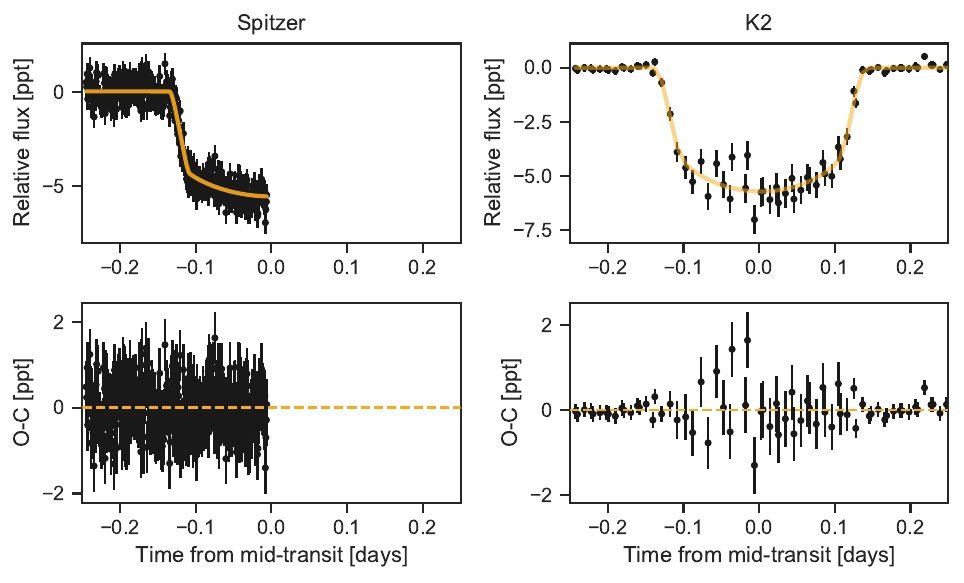}
    \caption{Top: Median model predictions and the 68\% percentile error bands for the (a) linear, (b) quadratic, and (c) quasi-periodic Gaussian process RV trend models. The data shown are from Keck/HIRES. The mean model residuals are plotted in the bottom panels. Bottom: Best-fit models for the \emph{Spitzer} (left) and \emph{K2} (right) photometry. The data and models are shown in the top panels, and the residuals in the bottom panels. The large scatter in-transit in the \emph{K2} data is likely due to starspot crossings.}
    \label{fig:rmfits}
\end{figure*}

\begin{figure*}
    \centering
    \includegraphics[width=\linewidth]{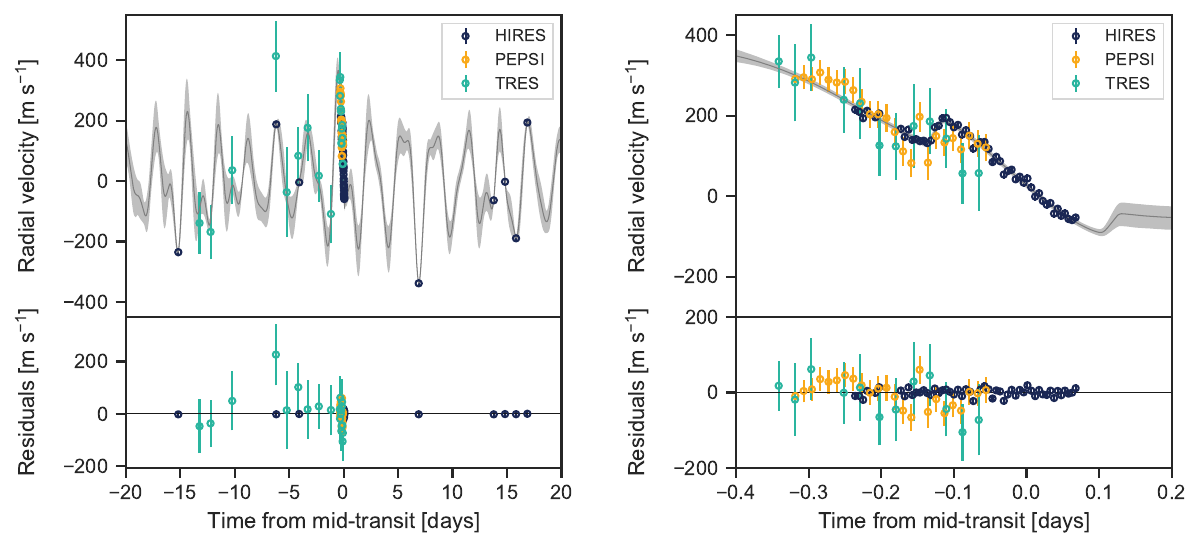}
    \caption{Mean model prediction and the 68\% percentile error band (grey) for the quasi-periodic Gaussian process + RM fit to the combined HIRES, PEPSI blue, and TRES datasets with a tight \prot prior. The data have been mean-subtracted and the mean model residuals are plotted in the lower panels. At left is the RV time series centered around the the RM observation night and at right is a zoom-in of the RM sequence.}
    \label{fig:gprmfit}
\end{figure*}

\begin{figure}
    \centering
    \includegraphics[width=\linewidth]{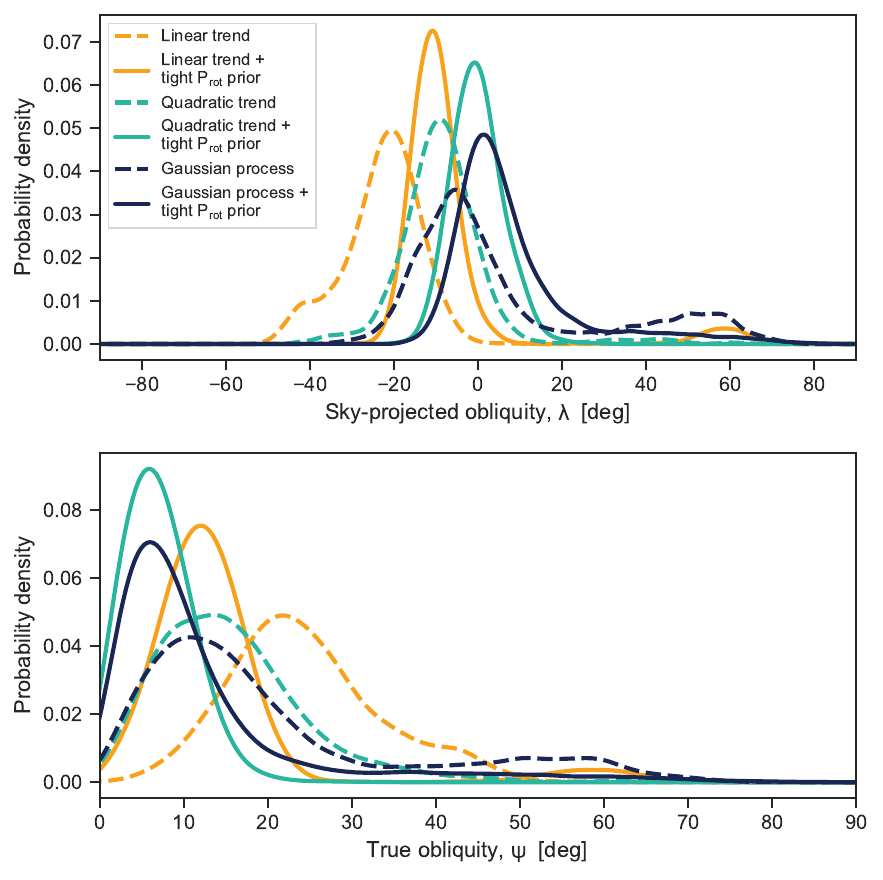}
    \caption{Comparison of posterior densities in sky-projected obliquity (top) and obliquity (bottom) for several of the RVRM models described in \S\ref{subsec:RVRM}. Posteriors shown are from fits to the HIRES data and have been smoothed using Gaussian kernel density estimation for clarity.}.
    \label{fig:posteriors}
\end{figure}

For a selection of models we show the mean model predictions and uncertainty bands in Figures~\ref{fig:rmfits} \& \ref{fig:gprmfit}. We summarize the inferred obliquities from these fits in Table~\ref{tab:rvrm-summary} and Figure~\ref{fig:posteriors}. Detailed MCMC summary statistics for the preferred model (the quasi-periodic GP with a tight \prot prior, using only the HIRES data) are presented in Table~\ref{tab:mcmc}, and we show the posterior distributions in Fig.~\ref{fig:corner_gp_protprio_hires}. The priors used in the RVRM analysis are summarized in Table~\ref{tab:rvrm-priors}, The data and code needed to reproduce our RVRM analysis are publicly available through GitHub.\footnote{\url{https://github.com/trevordavid/obliquity}}. The MCMC posteriors will be made publicly available through a Zenodo repository upon acceptance. 

The adopted trend model does have a significant effect upon the derived obliquity, as shown in Fig.~\ref{fig:posteriors}. Although the orbit is clearly prograde, the choice of trend model impacts the inferred value of the sky-projected spin-orbit misalignment $\lambda$ at the $\sim10^{\circ}-20^{\circ}$ level, and some of the distributions have tails to values of $|\lambda|$ as high as $\sim60^{\circ}$. 

We ultimately adopt the Gaussian process models as our preferred solution; this model most carefully accounts for the RV trend, accounting for all of the information that we have on the stellar RV variability over timescales of hours to weeks. Additionally, we adopt the tight $P_{\mathrm{rot}}$ prior as this best encompasses prior knowledge of the system, but note that using the looser prior instead does not qualitatively change the results. Although the residuals for the transit RVs in this model are slightly larger than for the quadratic model, the difference is small (9 cm s$^{-1}$).

\begin{deluxetable}{lccc}
\tabletypesize{\scriptsize}
\tablecolumns{4}
\tablewidth{0pt}
\tablecaption{Inferred obliquities from the RVRM analyses. \label{tab:rvrm-summary}}
\tablehead{
\colhead{Model} & 
\colhead{Instruments} & 
\colhead{$\lambda$} & 
\colhead{$\psi$} \\
\colhead{} & 
\colhead{} & 
\colhead{(deg)} & 
\colhead{(deg)}
}
\startdata
Linear trend & HIRES & -22$_{-8}^{+9}$ & 23$_{-10}^{+7}$ \\
Linear trend, tight \prot prior & HIRES & -10$_{-6}^{+4}$ & 12$_{-5}^{+5}$ \\
Quadratic trend & HIRES & -9$_{-7}^{+8}$ & 14$_{-9}^{+6}$ \\
Quadratic trend, tight \prot prior & HIRES & -1$_{-6}^{+5}$ & 6$_{-4}^{+3}$ \\
Quasi-periodic GP & HIRES & -2$_{-16}^{+11}$ & 15$_{-12}^{+8}$ \\
\textbf{Quasi-periodic GP, tight \prot prior} & \textbf{HIRES} & \textbf{4$_{-10}^{+7}$} & \textbf{8$_{-7}^{+4}$} \\
Quasi-periodic GP & HIRES, PEPSI, TRES & 8$_{-23}^{+20}$ & 16$_{-14}^{+10}$ \\
Quasi-periodic GP, tight \prot prior & HIRES, PEPSI, TRES & 10$_{-16}^{+12}$ & 12$_{-12}^{+8}$ \\
\enddata
\tablecomments{Values quoted are the medians and 68\% highest density intervals from the posteriors. Adopted values in bold.}
\end{deluxetable}

\subsection{Doppler Tomographic Analysis}
\label{sec:DT}

V1298 Tau is sufficiently rapidly rotating (\vsinistar$\sim24$ \kms) that we can spectroscopically resolve the rotationally broadened line profile. The RV Rossiter-McLaughlin effect arises from the perturbation to the line profile due to the missing light from the planetary transit; the centroid shift due to this perturbation results in the RV anomaly during transit that we detected in \S\ref{subsec:RVRM}. Due to V1298 Tau's rapid rotation, we can hope to directly measure this line profile perturbation in the time series line profiles. This method is typically known as Doppler tomography \citep[e.g.,][]{Albrecht07,CollierCameron10W33,Kepler13Ab}. Compared to the RM technique where we model the shifts in the line centroids, in Doppler tomography we instead directly model the line shape.

Our methodology for extracting the average line profiles from the time series spectra was largely the same as that used in \cite{Kepler13Ab,WASP33b,DT3,KELT21}. In short, we used the least squares deconvolution method \citep{Donati97}, fitting a model produced by convolving a line profile with a picket fence of $\delta$ functions at the wavelengths of the spectral lines to the data. We first fit for the depths of the individual spectral lines, and then for the line profile itself. We perform these fits to the data on an order-by-order basis; the final extracted line profile from a spectrum is the weighted average of these order-by-order line profiles, weighted based on the signal-to-noise and total equivalent width of lines in each order.

We used the full available bandwidth of PEPSI (4800-5441 \AA\ and 6278-7419 \AA) to extract the line profiles. 
For the HIRES data we extracted line profiles only from the blue and red chip data (covering 3646-4790 \AA\ and 6549-7984 \AA, respectively); the green chip data are heavily impacted by iodine absorption lines and therefore not usable for the Doppler tomographic analysis. Additionally, in order to increase the signal-to-noise, we binned together sets of three consecutive HIRES spectra before extracting the line profiles. We also do not consider the TRES spectra in this analysis because they are too low signal-to-noise.

\begin{figure}
    \centering
    \includegraphics[width=\columnwidth]{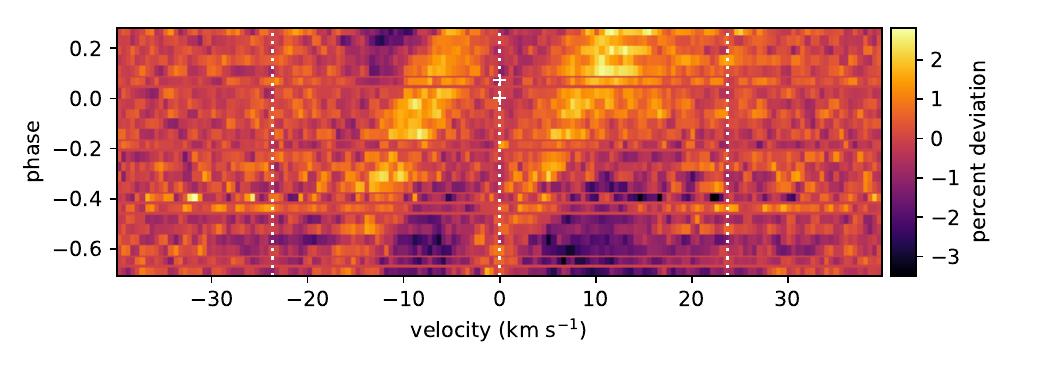} \\
    \includegraphics[width=\columnwidth]{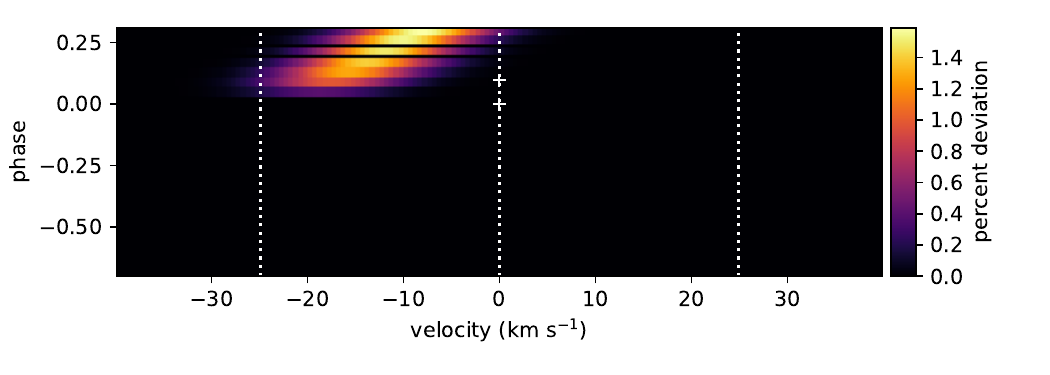} \\
    \includegraphics[width=\columnwidth]{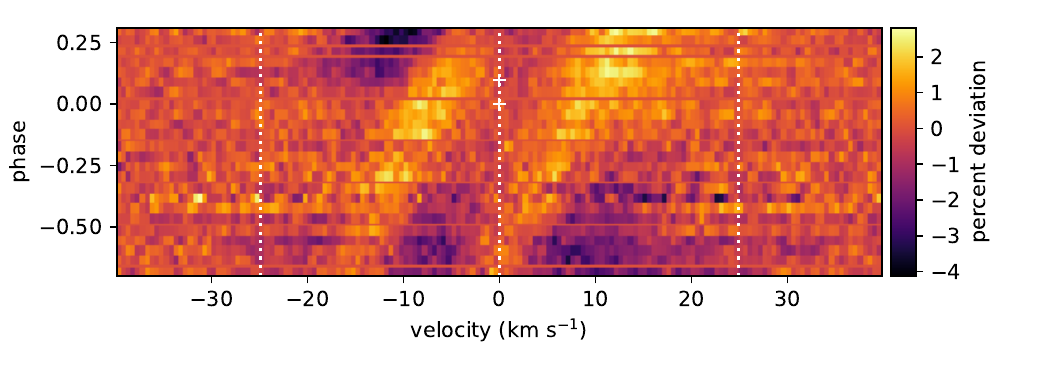} \\
    \includegraphics[width=\columnwidth]{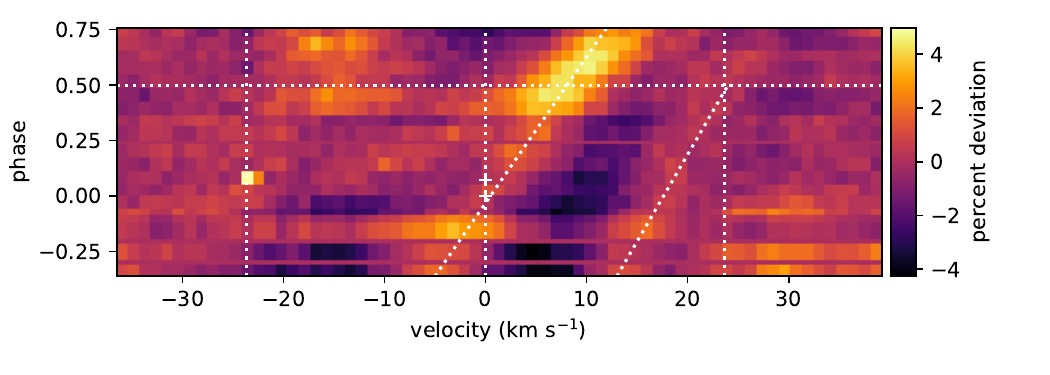}
    \caption{Doppler tomographic data from PEPSI (top three panels) and HIRES (bottom). All plots show the time series line profile residuals, i.e., each horizontal line shows the deviation of the line profile from the average line profile at that time. Time increases from bottom to top; units are such that ingress=0 and egress=1. The vertical dotted lines show $v=0, \pm$ \vsinistar, the horizontal dotted line the time of mid-transit, and the two small plus signs first and second contact. In the HIRES plot we also show two slanted dotted lines to guide the eye along the spot signatures, which are less obvious than in the PEPSI data.}
    \label{fig:dt}
\end{figure}

We show our time series line profile residuals in Fig.~\ref{fig:dt}.
The line profile residuals are dominated by two streaks corresponding to two spot complexes moving across the stellar surface during the observations. These are most obvious in the PEPSI data, but are also present in the HIRES data. These corroborate the spotted nature of the star as being the source of the trend seen in the RV data. The track of the planet is not obvious by eye.

Our efforts to model out the line profiles in the combined dataset and detect the planetary transit in the Doppler tomographic data have not yet born fruit. We defer a full quantitative analysis of these data to a future publication.

We can, however, analyze the spot complexes themselves. We have two clearly detected spot complexes. The distributions of spots on a stellar surface can inform knowledge of the magnetic geometry. The spot longitude $\phi$ can easily be measured from the line profiles as 
\begin{equation}
    \sin\phi=\frac{v}{\vsinistar}
\end{equation}
 where $\phi=0$ is defined to be the sub-observer longitude. 
 
 We measured the spot longitudes from each of the pre-transit PEPSI spectra by fitting a Gaussian profile to the spots in the time series line profile residuals. We fit the residuals from each spectrum separately, and take the mean and standard deviation of the spectrum where both spots were fit successfully as the measurement. We thus estimate a longitudinal separation between the two spots of $\Delta\phi=42^{\circ}\pm4^{\circ}.$ There is no apparent trend in $\Delta\phi$ over the course of the observations, which could be caused by differential rotation. The motion of the spots is consistent with the photometric \prot\ determined by \cite{David19b}. The time coverage of the observations is also too short to perform a full inversion to determine the overall surface spot distribution and determine the latitudes of the spots. 
 
 On the Sun, sunspots are often separated by $180^{\circ}$ of longitude \citep{BerdyuginaUsokin03}. Although it is certainly possible that the $45^{\circ}$ separation between the two spot complexes we see on V1298 Tau could be a coincidence, it could also hint at a non-solar-like dynamo. More observations of V1298 Tau would be required to test this hypothesis, to determine whether a $45^{\circ}$ separation is typical. We note, however, that \cite{Feinstein21} did not see any large spot complexes in their data, obtained a few months after our observations.
 
 In principle we could also measure the physical size of the spot complexes from the velocity space extension of the line profile perturbations. Even with the $R=120,000$ PEPSI data, however, the spot signatures are of similar size to the instrumental broadening profile and so we consider them to be unresolved. Nonetheless, the instrumental resolution of 2.5 \kms corresponds to a best latitude resolution of $\sim6^{\circ}$, which allows us to estimate an upper limit on the spot complex size of $<0.15$ \rsun.

\section{Discussion}
\label{sec:Discussion}

\begin{figure}
\centering
\includegraphics[width=\columnwidth]{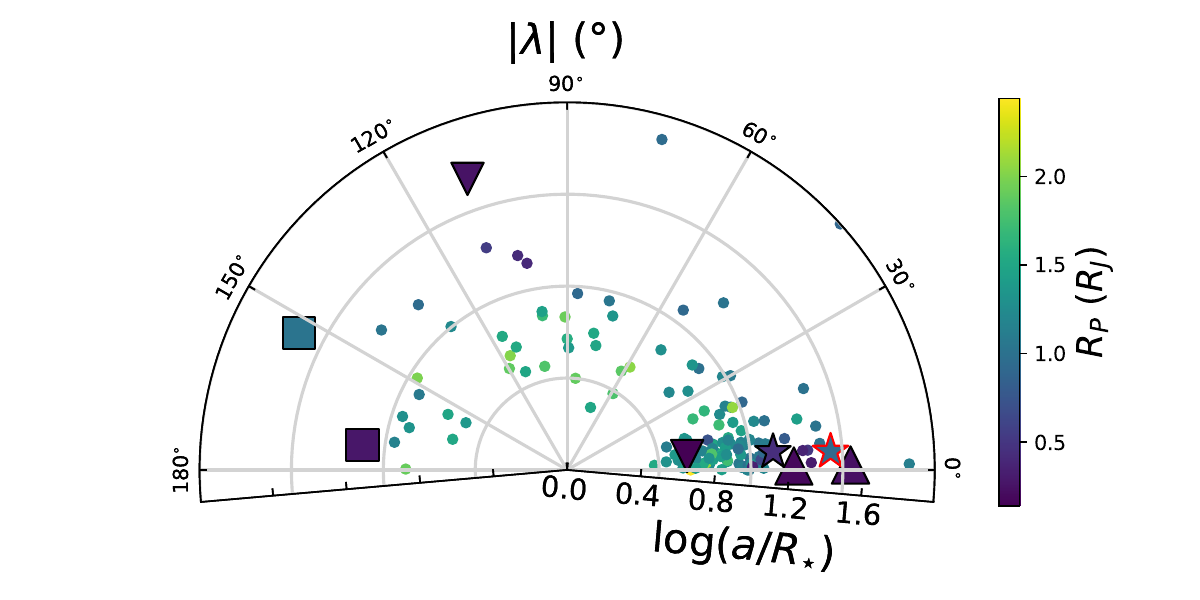}
\includegraphics[width=\columnwidth]{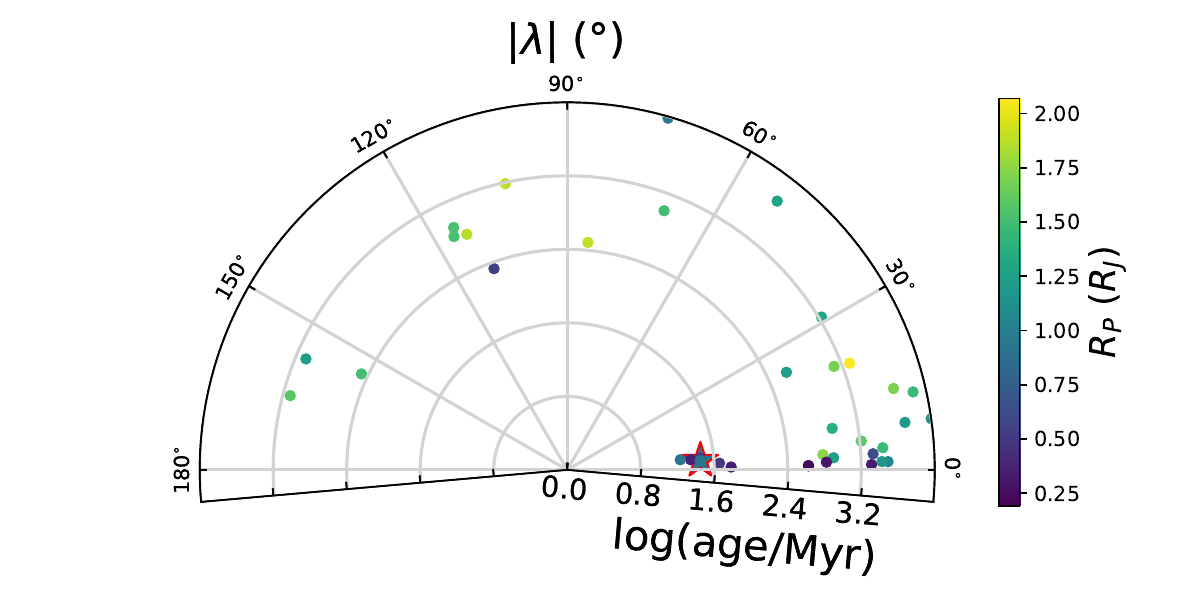}
\caption{Our measurement of the sky-projected spin-orbit misalignment of V1298 Tau b in context with measurements from the literature. Top: all published spin-orbit misalignments, shown as a function of the scaled semi-major axis $a/R_{\star}$. The color scale shows the planetary radius. We highlight V1298 Tau b as the red-bordered star, and also show as larger symbols the four systems with published spin-orbit misalignment measurements for more than one planet: V1298 Tau (stars), HD 63433 (upward triangles), K2-290 (squares), and HD 3167 (downward triangles). Note that K2-290 is consistent with coplanarity, as the uncertainty on the measurement of $\lambda$ for the inner planet is large \citep{Hjorth21}, while HD 3167 is highly non-coplanar \citep{Bourrier21}. Bottom: spin-orbit misalignments as a function of age; young planets are towards the center, older planets towards the edge. As has been noted by other recent works \citep[e.g.,][]{Heitzmann21}, all of the young planets ($<100$ Myr age) with published measurements are aligned. However, the number of such planets is small. Again, V1298 Tau b is highlighted.}
\label{fig:lambdapop}
\end{figure}

\planet joins a growing number of planets in young stellar associations with measured spin-orbit angles, including DS Tuc Ab \citep{Zhou20,Montet20,Benatti2021}, AU~Mic~b \citep{Hirano20, Martioli20, Palle20, Addison20}, HD~63433~b \citep{Mann20} and c \citep{Dai20}, HIP 67522~b \citep{Heitzmann21}, and V1298 Tau b's sister planet V1298 Tau c \citep{Feinstein21}. Spin-orbit angles have also been determined for several planets transiting young field stars, including Kepler-63~b \citep{SanchisOjeda13}, TOI-942~b \citep{Wirth21}, KELT-9~b \citep{Gaudi17, Ahlers20}, KELT-20~b \citep{Lund17}, and TOI-1431~b/MASCARA-5~b \citep{Addison21}. The aforementioned planets in young associations, TOI-942~b, and KELT-20~b have all been found to have aligned orbits, while Kepler-63~b, KELT-9~b, and TOI-1431~b/MASCARA-5~b are on polar orbits. We show this population in Fig.~\ref{fig:lambdapop}. Thus, it appears that at least some spin-orbit misalignments can arise in the first few hundred Myr. Furthermore, the recent discovery of two planets on retrograde orbits around a field star is challenging to explain with post-formation evolution \citep{Hjorth21}.

V1298 Tau~b, with an orbital period of 24 days, is also among the longest-period planets for which the spin-orbit angle has been measured, as can be seen in the top panel of Fig.~\ref{fig:lambdapop}.

The low obliquity of \planet suggests that the various theoretical mechanisms proposed to produce primordial spin-orbit misalignments were not important enough to produce a gross misalignment in the system. Notably, there is no known stellar companion to the V1298 Tau system \citep{David19}. We can not rule out mild misalignments like that of the solar system. Of course, it is still possible that those same mechanisms may be important for some fraction of planetary systems. 

There are now six planets in five systems with ages between 15 and 60 Myr with spin-orbit angle measurements, all of which are aligned. Although these are still small-number statistics, and a quantitative assessment will have to wait for the accumulation of more measurements, it appears that the majority of young systems are well aligned. If this trend is borne out by observations of more young planets, it could suggest that misaligned orbits are typically generated by dynamical mechanisms with longer timescales like secular chaos or the Kozai-Lidov effect.

It is unclear, however, how the young planet population relates to the older planets that account for most of the spin-orbit misalignment measurements to date. Most of the young planets do not have measured masses, and only HIP 67522~b and V1298 Tau~b are near Jupiter-radius; meanwhile, two of the five young systems have multiple transiting planets. Furthermore, V1298 Tau hosts two planets with masses close to that of Jupiter with periods of 40 days or less \cite{SuarezMascareno21}, again setting it apart for the bulk of the spin-orbit misalignment sample. Only the HIP 67522 b system could resemble the hot Jupiters that make up most of the field-age sample. Conversely, only a few older small and multi-planet systems have spin-orbit measurements. The multi-planet systems are generally well-aligned, with a few exceptions \citep{Huber13,Dalal19,Hjorth21}, while lonely hot Neptunes may be more commonly misaligned \citep{Winn10HAT11,Bourrier18,Rubenzahl21}. The aligned young planets may simply belong to a population that is typically aligned at all ages. 
More spin-orbit measurements for both young planets and older small planets, as well as mass measurements for the young planets, are needed to draw firm conclusions.

\subsection{Mutual Inclination and System Architecture}

\begin{figure}
\centering
\includegraphics[width=\columnwidth]{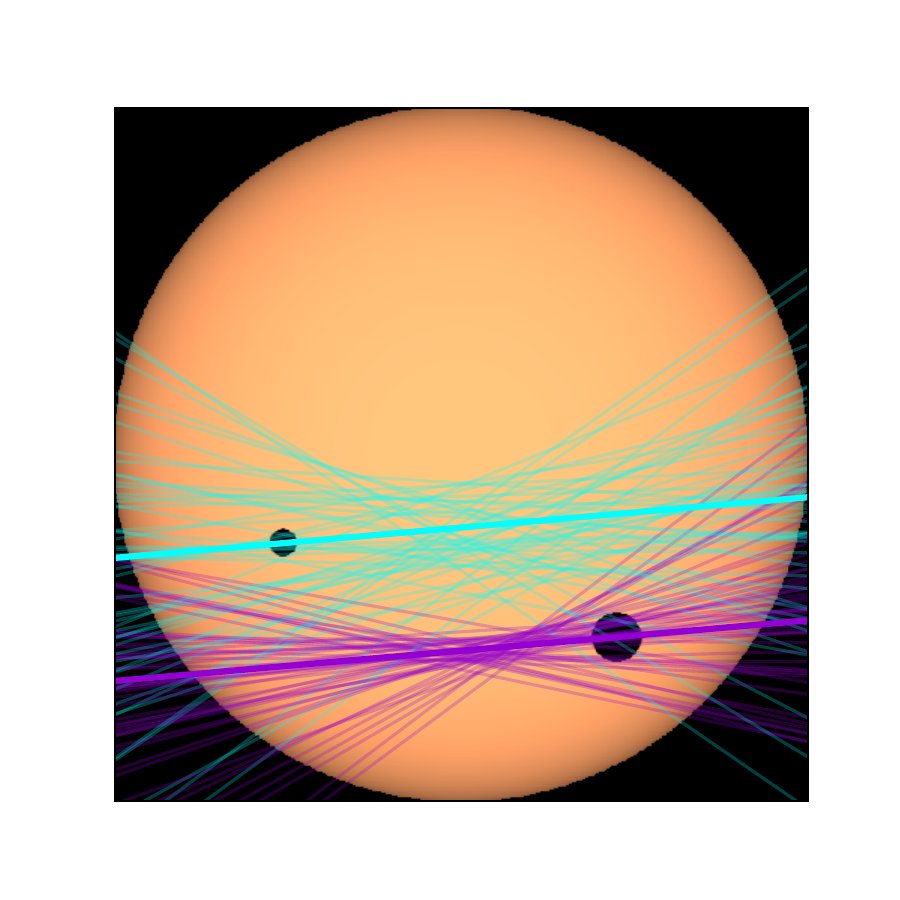}
\caption{Schematic of the transit geometry of V1298 Tau b (large disk, purple) and c (small disk, cyan). The thick lines show the best-fit transit chords for each planet from this work and \cite{Feinstein21}, respectively, while the thin lines show 50 random draws from the posterior distributions; note that we only consider uncertainties in $\lambda$ and $b$, which are the parameters that directly govern the path of the planet across the stellar disk.}
\label{fig:transitgeometry}
\end{figure}

The recent measurement of the spin-orbit angle for V1298 Tau c by \cite{Feinstein21} presents an opportunity to constrain the mutual inclination of the two planets. They found that planet c is consistent with an aligned orbit, with $\lambda_c=5^{\circ}\pm15^{\circ}$. We show a schematic of the transit geometry of the system in Fig.~\ref{fig:transitgeometry}.

The mutual inclination between two planets with known values of $i$ and $\lambda$ can be expressed as:
\begin{equation} \label{eq:1}
    \cos i_{\mathrm{mut}} = \sin i_1 \sin i_2 \cos (\lambda_1 - \lambda_2) + \cos i_1 \cos i_2
\end{equation}
where $i_{\mathrm{mut}}$ is the mutual inclination between the two planets. We present the derivation of Equation~\ref{eq:1} in Appendix~\ref{sec:math-appendix}.

We find a mutual inclination between V1298 Tau b and c of \imutfinal. Although this is not particularly constraining, this confirms that these two planets are consistent with a coplanar configuration. This is expected given the presence of four transiting planets in this system \citep{David19b}.

Future observations of both planets, as well as the other planets orbiting V1298 Tau, to more precisely measure the spin-orbit misalignment could enable meaningful constraints on the mutual inclinations and overall architecture of this system.

\section{Conclusions}
\label{sec:Conclusions}

We summarize our key findings below:
\begin{enumerate}
    \item The RM effect is clearly detected during a partial transit of V1298 Tau b, indicating a prograde orbit and a likely low orbital obliquity for the planet. V1298 Tau b is one of the youngest and coolest exoplanets for which the orbital obliquity has been constrained.
    \item From modeling of the partial RM curve we find the inferred sky-projected obliquity of V1298 Tau, with respect to the orbit of planet b, is sensitive to the underlying RV trend. A range of obliquities are compatible with the data but, for most assumed models of the RV trend, sampling from the posterior distributions indicates that the highest posterior density is at low or moderate obliquities ($\left | \lambda \right | \lesssim$ 20$^\circ$) and the data are compatible with an obliquity of zero. For our most realistic model of the stellar RV trend we find a sky-projected obliquity of \lambadfinal and a obliquity of \psifinal. 
    \item Combining our obliquity constraints with those of the interior planet V1298 Tau c published in \citet{Feinstein21} we constrained the mutual inclination of the two planets to be \imutfinal. The available data are thus consistent with a coplanar configuration and a low stellar obliquity, as expected from formation within a circumstellar disk.
    
    \item Using newly available data from Gaia DR3, we revise the age of V1298 Tau to \agefinal.
    
    \item At the epoch of our spectroscopic observations, we find the existence of two spot complexes on V1298 Tau separated by $\Delta\phi=42^{\circ}\pm4^{\circ}$ of longitude.
    
    \item From our joint fits of the RVs and photometry we derive a new, more precise transit ephemeris for V1298 Tau~b:
    \begin{equation*}
        T_\mathrm{mid} = 2458781.0835 \pm 0.0013
    \end{equation*}
     \begin{equation*}
        P = 24.141341 \pm 0.000023 \mathrm{\, days.}
    \end{equation*}
    \item RVs derived from the PEPSI spectrograph clearly demonstrate the chromaticity of the underlying RV trend, which we attribute to rotation of the inhomogeneous stellar surface. We find RV slopes of $\sim$45 m~s$^{-1}$~hour$^{-1}$ in the blue arm (4800-5441\AA) and $\sim$95 m~s$^{-1}$~hour$^{-1}$ in the red arm (6278-7419\AA). It is notable that a steeper trend is observed at redder optical wavelengths, counter to expectations from how stellar activity scales with wavelength \citep[e.g.][]{Tran2021}. One possible explanation is that the PEPSI red radial velocities are derived from a larger set of spectral lines that are particularly prominent in starspots, thereby biasing the radial velocities more strongly in the red compared to the blue data.

    \item We demonstrate that an intra-night RV precision of 6-7~\ms is achievable on V1298 Tau, a highly active pre-main sequence star, using Keck/HIRES with the iodine cell and the California Planet Search RV pipeline. 
\end{enumerate}

The same time series spectra through the course of a transit used for spin-orbit misalignment measurements are also often usable for transmission spectroscopy to investigate the planetary atmosphere and exosphere. We analyze the time-variable H$\alpha$ line profile of V1298 Tau in our data in a companion paper (David et al. in prep).

\vspace{12pt}

Thanks to Adina Feinstein, Ben Montet, and Elisabeth Newton for useful discussions. We thank the other observers who contributed some of our observations or helped with the planning of the observations: David Latham, Samuel Quinn, and Andrew Howard. Thanks to Josh Walawender for supporting our Keck observations.

This work was supported by a NASA Keck PI Data Award through JPL RSA 1634873. P. D. is supported by a National Science Foundation (NSF) Astronomy and Astrophysics Postdoctoral Fellowship under award AST-1903811.

Some of the data presented herein were obtained at the W. M. Keck Observatory, which is operated as a scientific partnership among the California Institute of Technology, the University of California and the National Aeronautics and Space Administration. The Observatory was made possible by the generous financial support of the W. M. Keck Foundation. 
This research has made use of the Keck Observatory Archive (KOA), which is operated by the W. M. Keck Observatory and the NASA Exoplanet Science Institute (NExScI), under contract with the National Aeronautics and Space Administration. 
The LBT is an international collaboration among institutions in the United States, Italy and Germany. The LBT Corporation partners are: The Ohio State University; LBT Beteiligungsgesellschaft, Germany, representing the Max Planck Society, the Astrophysical Institute Potsdam, and Heidelberg University; The University of Arizona on behalf of the Arizona university system; Istituto Nazionale di Astrofisica, Italy; The Research Corporation, on behalf of The University of Notre Dame, University of Minnesota and University of Virginia.
%This work makes use of observations from the LCOGT network. 
This paper includes data collected by the K2 mission. Funding for the K2 mission is provided by the NASA Science Mission directorate. This work is based in part on observations made with the Spitzer Space Telescope, which was operated by the Jet Propulsion Laboratory, California Institute of Technology under a contract with NASA.
This work has made use of data from the European Space Agency (ESA) mission
{\it Gaia} (\url{https://www.cosmos.esa.int/gaia}), processed by the {\it Gaia}
Data Processing and Analysis Consortium (DPAC,
\url{https://www.cosmos.esa.int/web/gaia/dpac/consortium}). Funding for the DPAC
has been provided by national institutions, in particular the institutions
participating in the {\it Gaia} Multilateral Agreement.

The authors wish to recognize and acknowledge the very significant cultural role and reverence that the summit of Maunakea has always had within the indigenous Hawaiian community.  We are most fortunate to have the opportunity to conduct observations from this mountain.

\facility{Keck:I (HIRES), LBT (PEPSI), FLWO1.5m (TRES), \textit{Kepler}, \textit{Spitzer}}

\software{\texttt{astropy} \citep{exoplanet:astropy13, exoplanet:astropy18},
          \texttt{celerite2} \citep{celerite2:foremanmackey17, celerite2:foremanmackey18},
          \texttt{everest} \citep{Luger16,Luger18},
          \texttt{exoplanet} \citep{exoplanet:exoplanet, exoplanet:joss},
          \texttt{jupyter} \citep{jupyter},
          \texttt{matplotlib} \citep{matplotlib},
          \texttt{misttborn} \citep{DT3},
          \texttt{numpy} \citep{numpy},
          \texttt{pandas} \citep{pandas-soft, pandas-proc},
          \texttt{pymc3} \citep{exoplanet:pymc3},
          \texttt{seaborn} \citep{seaborn},
          \texttt{starry} \citep{exoplanet:luger18},
          \texttt{theano} \citep{exoplanet:theano}
          }

\appendix 

\section{Derivation of the Mutual Inclination Formula}
\label{sec:math-appendix}

Here we present the derivation of the formula giving the mutual inclination $i_{\mathrm{mut}}$ between the orbits of two planets 1, 2 with orbital inclinations $i_1$, $i_2$ and sky-projected spin-orbit misalignments $\lambda_1$, $\lambda_2$.

We define a coordinate system $xyz$ such that the $z$ axis is along the line of sight and the $y$ axis is parallel to the projected stellar spin axis. Let $\mathbf{\hat{y}_n}$ be a unit vector along the orbital angular momentum of planet $n$. %To express this vector in the observer's frame, we rotate the coordinate system by $\pi/2-i_n$ about the $z$ axis, and then by $\lambda_n$ about the $x$ axis:
Let us begin with a system where the orbital plane is parallel to the line of sight and with zero obliquity. For an arbitrary system configuration, we need to rotate $\mathbf{\hat{y}_n}$ to the actual orientation of the planetary orbit. We can accomplish this rotation by first rotating the vector by $\pi/2-i_n$ about the $x$ axis to account for the mutual inclination, and then by $\lambda_n$ about the $z$ axis to account for the projected obliquity:
\begin{equation}
    \mathbf{\hat{y}_n'} = \mathit{R}_z(-\lambda_n) \mathit{R}_x(\pi/2-i_n) \mathbf{\hat{y}_n} 
\end{equation}

where $\mathit{R}_z$ and $\mathit{R}_x$ are rotation matrices about the $z$ and $x$ axes, respectively.
The angle between the vectors for the two planets is then:
\begin{equation}
    \cos i_{\mathrm{mut}} = \mathbf{\hat{y}_1}'^T \cdot \mathbf{\hat{y}_2'}
\end{equation}

Working through these equations then yields
\begin{equation}
    \cos i_{\mathrm{mut}} = \sin i_1 \sin i_2 \cos (\lambda_1 - \lambda_2) + \cos i_1 \cos i_2
\end{equation}

In the case that both planetary orbits are edge-on ($i_1 = i_2 = 90^{\circ}$), the mutual inclination is simply the difference in the spin-orbit angles:
\begin{equation}
    \cos i_{\mathrm{mut}} = \cos (\lambda_1 - \lambda_2)
\end{equation}

In the case that both projected spin-orbit angles are identical, the mutual inclination is simply the difference in inclinations:
\begin{equation}
    \cos i_{\mathrm{mut}} = \cos (i_1 - i_2)
\end{equation}

\section{RVRM Tables}

We show the priors used in our MCMC analysis of the RV RM data (described in \S\ref{subsec:RVRM}) in Table~\ref{tab:rvrm-priors}, and list the results and summary statistics from this MCMC in Table~\ref{tab:mcmc}.

\begin{deluxetable}{lcc}
\tabletypesize{\scriptsize}
\tablecolumns{3}
\tablewidth{0pt}
\tablecaption{Priors used in the RVRM analyses. \label{tab:rvrm-priors}}
\tablehead{
\colhead{Parameter} & 
\colhead{Prior} & 
\colhead{Models}
}
\startdata
$M_\star$ (\msun) & $\mathcal{N}$(1.10, 0.05) & L/Q/GP\\
$R_\star$ (\rsun) & $\mathcal{N}$(1.305, 0.07) & L/Q/GP\\
P (d)           & $\mathcal{N}$(24.141445, 0.000056)& L/Q/GP\\
T0 - offset (d)       & $\mathcal{N}$(0, 0.1)& L/Q/GP\\
$R_\mathrm{P}/R_\star$     & $\mathcal{U}$(0.0,0.2)& L/Q/GP\\
$\sigma_\mathrm{Spitzer}$ (ppt) & $\mathcal{N}$(0,10) & L/Q/GP\\
$\sigma_\mathrm{Kepler}$  (ppt) & $\mathcal{N}$(0,10) & L/Q/GP\\
ln jitter$_{\mathrm{Spitzer}}$ (nat) & $\mathcal{U}$(-10,2) & L/Q/GP\\
ln jitter$_{\mathrm{Kepler}}$ (nat) & $\mathcal{U}$(-10,2) & L/Q/GP\\
$v\sin(i_\star)$~(\kms) & $\mathcal{N}$(24.9, 0.2) & L/Q/GP\\
$\cos(i_\star)$ & $\mathcal{U}$(0,1) & L/Q/GP\\
$P_\mathrm{rot}$ (d) & $\mathcal{N}$(2.87, 0.022 or 0.2)$^*$ & L/Q/GP\\
$\lambda$ (rad) & $\mathcal{U}$(-$\pi$, $\pi$) & L/Q/GP\\
$c_0$ (\ms) & $\mathcal{N}$(0, 10$^5$) & L/Q \\
$c_1$ (\ms d$^{-1}$) & $\mathcal{N}$(0, 10$^5$) & L/Q\\
$c_2$ (\ms d$^{-2}$) & $\mathcal{N}$(0, 10$^5$) & Q\\
ln jitter$_{\mathrm{HIRES}}$ (nat) & $\mathcal{N}$(ln($\sigma_{\mathrm{HIRES}}$), 5) & GP\\
ln jitter$_{\mathrm{TRES}}$ (nat) & $\mathcal{N}$(ln($\sigma_{\mathrm{TRES}}$), 5) & GP\\
ln jitter$_{\mathrm{PEPSI}}$ (nat) & $\mathcal{N}$(ln($\sigma_{\mathrm{PEPSI}}$), 5) & GP\\
$\mu_{\mathrm{HIRES}}$ (\ms) & $\mathcal{N}$(0, 50) & GP\\
$\mu_{\mathrm{TRES}}$ (\ms) & $\mathcal{U}$(-1000, 1000) & GP\\
$\mu_{\mathrm{PEPSI}}$ (\ms) & $\mathcal{N}$(-1000, 1000) & GP\\
ln $\sigma_{\mathrm{rot}}$ (nat) & $\mathcal{N}$(ln($\sigma_{\mathrm{RV}}$), 10) & GP\\
ln $Q_0$ (nat) & $\mathcal{N}$(1, 10) & GP\\
ln $\Delta Q$ (nat) & $\mathcal{N}$(2, 10) & GP\\
f & $\mathcal{U}$(0.1, 1) & GP\\
\enddata
\tablecomments{$\mathcal{N}(\mu, \sigma)$ denotes a normal prior with mean $\mu$, and standard deviation $\sigma$. $\mathcal{U}(a, b)$ denotes a uniform prior with lower limit $a$ and upper limit $b$. Impact parameter was sampled jointly with $R_\mathrm{P}$/$R_\star$ using the \citet{Espinoza2018} formalism. Limb darkening coefficients in the Kepler and Spitzer bandpasses were sampled from the uniformative priors suggested by \citet{exoplanet:kipping13}. L: Linear model, Q: quadratic model, GP: quasi-periodic Gaussian process. $^*$The width of the \prot prior is given for the tight and loose priors, respectively.}
\end{deluxetable}

\begin{deluxetable*}{lrrrrr}
\tabletypesize{\footnotesize}
\tablecolumns{6}
\tablewidth{0pt}
\tablecaption{MCMC summary statistics for the quasi-periodic GP model with a tight \prot prior (HIRES data only) \label{tab:mcmc}}
\tablehead{
\colhead{Parameter} &
\colhead{Mean} &
\colhead{Std. dev.} &
\colhead{HDI (3\%)} &
\colhead{HDI (97\%)} & 
\colhead{$\hat{R}$}
}
\startdata
 Period, $P$ (d)                                  &  24.141341 &   0.000023 &  24.141298 &   24.141385  &  1.0000 \\
 Time of mid-transit - offset, $T_0$ (d)                        &  -0.0054   &   0.0013   &  -0.0080   &   -0.0031    &  0.9998 \\
 ln $\sigma_\mathrm{rot}$                         &   5.53     &   0.53     &   4.57     &    6.54      &  1.0009 \\
 ln $Q_0$                                         &   7.7      &   6.5      &  -1.8      &   20.0       &  1.0001 \\
 ln $\Delta Q$                                    &   2.0      &   9.9      & -18.1      &   19.4       &  0.9998 \\
 \mstar (\msun)                                   &   1.135    &   0.047    &   1.050    &    1.228     &  1.0000 \\
 \rstar (\rsun)                                   &   1.407    &   0.018    &   1.375    &    1.441     &  1.0006 \\
 Radius ratio, $R_P/R_\star$                      &   0.071998 &   0.00062  &   0.070847 &    0.073131  &  1.0000 \\
 Impact parameter, $b$                            &   0.564    &   0.026    &   0.515    &    0.613     &  1.0000 \\
 Limb darkening parameter $u_0$                   &   0.54     &   0.23     &   0.09     &    0.93      &  1.0003 \\
 Limb darkening parameter $u_1$                   &  -0.03     &   0.27     &  -0.47     &    0.47      &  1.0017 \\
 Limb darkening parameter $u_\mathrm{Spitzer,1}$  &   0.56     &   0.14     &   0.27     &    0.79      &  1.0001 \\
 Limb darkening parameter $u_\mathrm{Spitzer,2}$  &  -0.15     &   0.18     &  -0.40     &    0.20      &  1.0003 \\
 ln jitter$_\mathrm{Spitzer}$                     &  -5.9      &   2.3      &  -9.8      &   -2.2       &  1.0006 \\
 $v$sin($i_\star$) (\kms)                                    &  24.77     &   0.19     &  24.40     &   25.12      &  1.0010 \\
 cos($i_\star$)                                   &   0.085    &   0.062    &  0.000015  &   0.196      &  1.0014 \\
 sin($i_\star$)                                   &   0.9944   &   0.0072   &   0.9805   &   1.0000     &  1.0014 \\
 $i_\star$ (deg)                                  &  85.1      &   3.6      &  78.7      &   90.0       &  1.0014 \\
 $v_\mathrm{eq}$ (\kms)                           &  24.91     &   0.26     &  24.44     &   25.41      &  1.0005 \\
 $a/R_\star$                                      &  26.06     &   0.46     &  25.14     &   26.87      &  0.9998 \\
 cos($i_\mathrm{pl}$)                             &   0.0217   &   0.0013   &   0.0190   &    0.0241    &  0.9998 \\
 $i_\mathrm{pl}$ (deg)                            &  88.759    &   0.077    &  88.620    &   88.911     &  0.9998 \\
 $\lambda$ (rad)                                  &   0.14     &   0.27     &  -0.23     &    0.74      &  1.0002 \\
 $\lambda$ (deg)                                  &   8        &  15        & -13        &   42         &  1.0002 \\
 $f$                                              &   0.57     &   0.25     &   0.17     &    1.0       &  0.9999 \\
 \enddata
 \tablecomments{HDI: Highest Density Interval. $\hat{R}$: Gelman-Rubin statistic. The reference time $T_0$ indicates the time of mid-transit relative to the date BJD = 2458781.089056.}
 \end{deluxetable*}

\section{Corner Plots}

In this Appendix we present the corner plots from our fits to determine the age of V1298 Tau (\S\ref{sec:age}) and measure the spin-orbit alignment of the planet (\S\ref{subsec:RVRM}).

\begin{figure}
\centering
\includegraphics[width=\columnwidth]{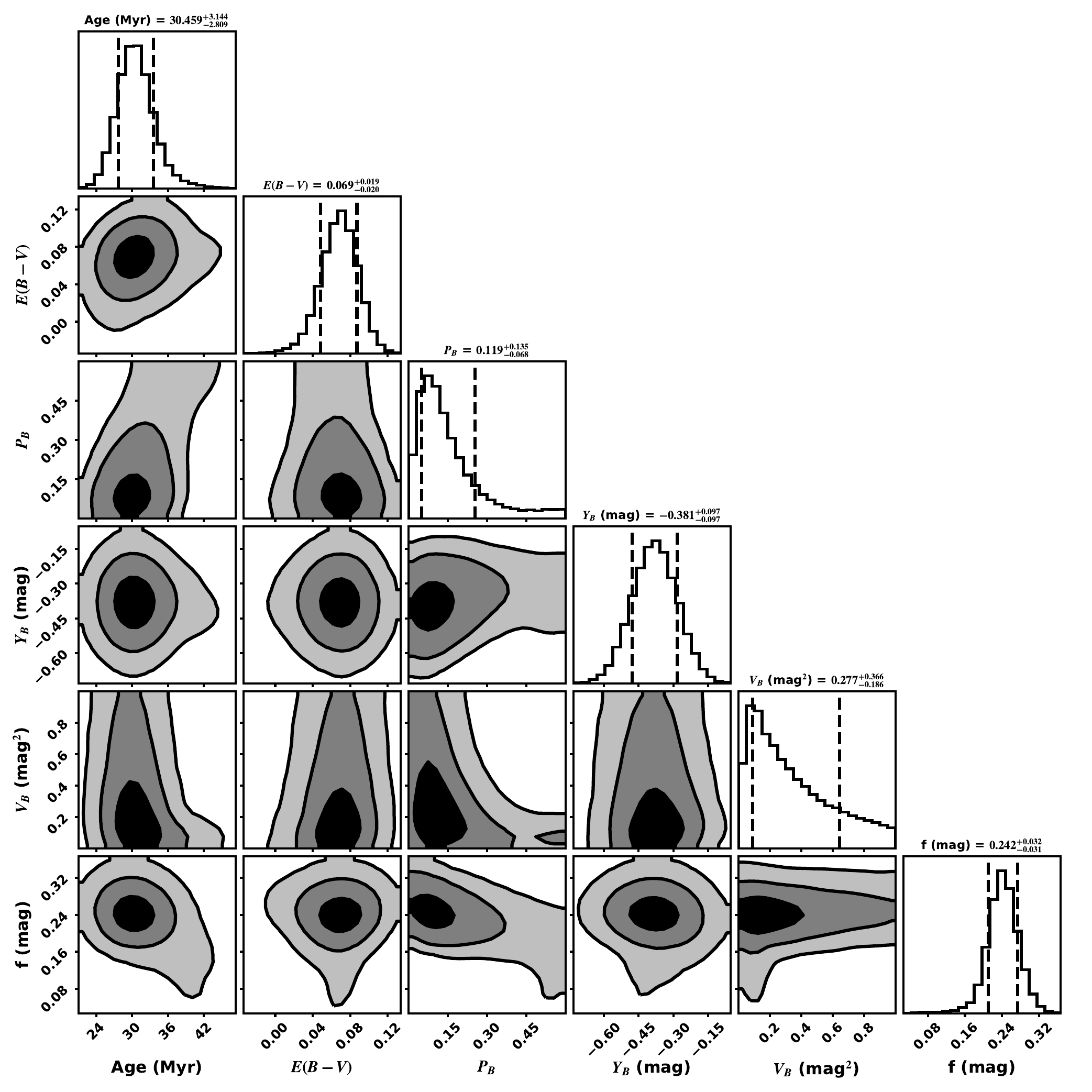}
\caption{Corner plot from the MCMC for fitting the age of V1298 Tau. We show all six fit parameters. $P_B$, $V_B$, and $Y_B$ are parameters that describe the outlier population, age and $E(B-V)$ describe the primary population, and $f$ applies to all stars in the sample (treated as a missing uncertainty). Similar fits using the DSEP magnetic models and with modified sample selection yield similar ages and reddening, but different outlier population parameters.
}
\label{fig:corner-cmd}
\end{figure}

 \begin{figure*}
    \centering
    \includegraphics[width=\linewidth]{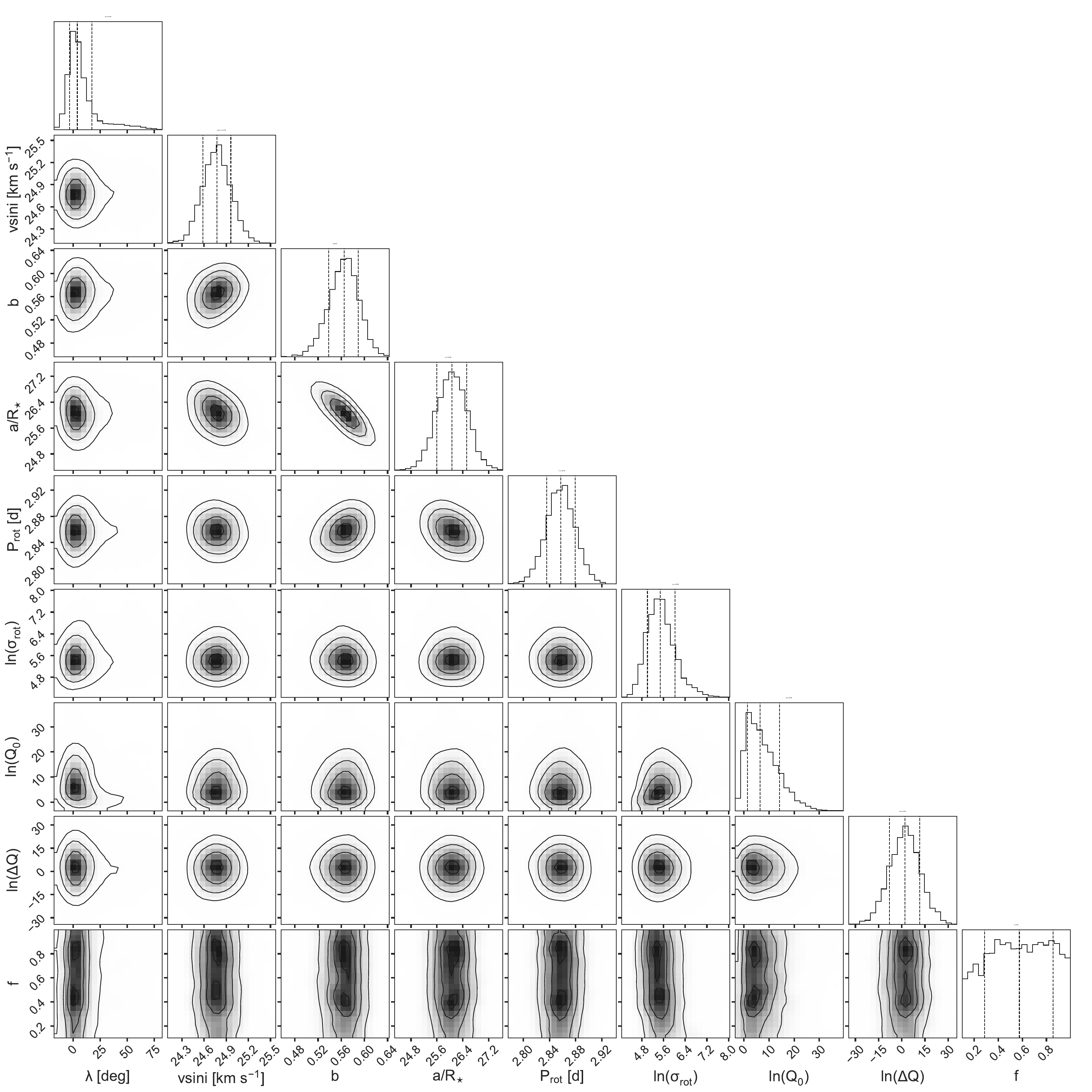}
    \caption{Posterior parameter distributions for the Rossiter-McLaughlin fit with the quasi-periodic Gaussian process + RM model with a tight \prot prior (for the HIRES data only). This is our adopted model.}
    \label{fig:corner_gp_protprio_hires}
\end{figure*}

\bibliography{bibmaster}{}

\end{document}